\documentclass[11pt]{article}
\usepackage{blois,epsfig}

\def\lesssim{\mathrel{\hbox{\rlap{\hbox{\lower4pt\hbox{$\sim$}}}\hbox{$<$}}}}
\def\gtrsim{\mathrel{\hbox{\rlap{\hbox{\lower4pt\hbox{$\sim$}}}\hbox{$>$}}}}

\def\arcdeg{\hbox{$^\circ$}}

\newcommand{\mm}[1]{\mbox{$#1$}}
\newcommand{\unit}[1]{\ifmmode \:\mbox{\rm #1}\else \mbox{#1}\fi}
\newcommand{\sbr}[1]{\mbox{$_{\rm #1}$}}

\newcommand{\mone}{\mm{^{-1}}}

\newcommand{\kms}{\unit{km~s\mone}}

\newcommand{\mpc}{\unit{Mpc}}

\newcommand{\hmpc}{\mm{h\mone}\mpc}

\newcommand{\lb}[2]{\mm{l = #1\arcdeg}, \mm{b = #2\arcdeg}}

\newcommand{\eqref}[1]{equation~(\ref{eq:#1})}

\newcommand{\aj}{\emph{AJ}}
\newcommand{\apj}{\emph{ApJ}}

\newcommand{\apjl}{\emph{ApJL}}
\newcommand{\mnras}{\emph{MNRAS}}

\begin{document}
\vspace*{4cm} 

\title{The Consistency of Cosmic Flows on 100 \hmpc\ Scales}

\author{M. J. Hudson}

\address{Department of Physics, University of Waterloo}

\maketitle

\section{Introduction}

Dark matter on 100 \hmpc\ scales is accessible to only a couple of
techniques. Cosmic Microwave Background (CMB) fluctuations probe dark
matter on these scales at $z \sim 1100$, but the anisotropy spectrum
is sensitive to other parameters as well as the dark matter power
spectrum.  Peculiar velocities offer a complementary approach, and
have the advantage that the dark matter distribution can be compared
directly to the distribution of galaxies in the nearby Universe. The
bulk flow statistic measures the mean motion of a sample with respect
to the CMB, and thus gives an indication of the level of mass density
fluctuations on scales larger than the sample size. Recent large-scale
peculiar surveys have measured bulk flows which, \emph{at face value},
appear to be in conflict. The purpose of this paper is to quantify the
effect of sparse sampling on the bulk flow statistic and to determine
whether recent results are consistent.

\section{Consistency of large-scale peculiar velocity surveys}

The SMAC cluster sample (Hudson \emph{et al.}\,\cite{smac}), with a
depth of $\sim~12000 \kms$, has a bulk velocity of $\sim 600 \kms$,
with respect to the Cosmic Microwave Background (CMB) frame. Some
other surveys (Willick\cite{w}; Lauer \& Postman\cite{lp}, hereafter
LP) have also yielded large bulk motions on similarly large scales.
However, Dale \emph{et al.}\,\cite{sc} (hereafter SC) found rather small
bulk motions on similar scales.  The EFAR survey (Colless et
al.\cite{efar}) was not designed to measure bulk flows but rather to
measure peculiar velocities near two distant superclusters.  As result
its sky coverage is very non-uniform and it is less suitable for bulk
flows.  We discuss the implications of non-uniform sampling below.

We have recently measured the bulk flow from the SNIa data of Tonry
\emph{et al}\cite{to}.  This sample contains a large number of objects
nearby ($R<6000 \kms$), where the bulk flow is known to be in the
range 300 -- 500 \kms.  To assess the bulk flow on very large scales,
beyond local attractors such as the ``Great Attractor'', we limit the
sample to the distant SNe with $6000 \kms < d < 30000 \kms$. We also
exclude SNe with extinction $A\sbr{V} > 1$.  This SN sample yields a
bulk flow of $610\pm200 \kms$ toward \lb{311\pm20}{8\pm15}, consistent
with with the bulk flows from the SMAC and Willick samples.

To address the consistency of cosmic flows, we have reanalyzed in a
consistent way the large-scale peculiar velocity samples discussed.
The results are given in Table 1.  The measurement errors are due to
peculiar velocity errors only; these are the values typically quoted
when reporting their bulk flow results. It is important to note that
these are accurate estimates of the bulk flow of the sparse peculiar
velocity samples, but do not necessarily reflect the error in the bulk
flow of the \emph{volume}. Based on these errors alone, there appears
to be conflict between some of the surveys (e.g. SC vs SMAC).

To calculate the bulk flow of a volume, one must be aware that
small-scale (``internal'') flows in a sparse sample do not completely
cancel, and will act as an extra source of noise.  In order to account
for this aliasing effect, it is necessary to have some idea of the
expected level of the internal flows. The statistical effect can be
calculated exactly if the power spectrum of mass fluctuations is
known.  Here we outline the main steps of the analysis; a more
detailed discussion is given in Hudson \emph{et al.}\,\cite{smac} (see
also Colless \emph{et al.}\,\cite{efar})

For each survey, we calculate the window functions for each Cartesian
component of the bulk flow (following Kaiser\cite{ka}).  The
contributions to the bulk flow statistic come from a wide range of
scales, with significant contributions from scales as small as
$\lambda \sim 30 \hmpc\ (k \sim 0.2)$. To assess the consistency of a
given survey with a cosmological model, we compute a total covariance
matrix ${\bf C} = {\bf C}\sbr{cos} + {\bf C}\sbr{pv}$, where the
subscript ``cos'' denotes the cosmic variance part and ``pv'' denotes
the peculiar velocity errors.  To compare two surveys, we generalize
this method. We calculate the difference in the sample bulk flows and
its total covariance, including the covariance due to random peculiar
velocities and the sampling covariance. The latter allows for the fact
that two sparse surveys do not trace the same volumes. For further
details of this approach, see Watkins \& Feldman\cite{wf}.

\begin{table}
\caption{Bulk flows and consistency for large-scale
surveys}\label{tab:bf}
\begin{minipage}[t]{\textwidth}
\begin{center}
\begin{tabular}{|llrrrrrrrr|}
\hline 
Survey & 
Method & 
$N$ & 
Depth & 
\multicolumn{1}{c}{\bf V} & 
\multicolumn{1}{c}{$l$} & 
\multicolumn{1}{c}{$b$} &
\multicolumn{1}{c}{\parbox{0.3in}{\small Meas.\ error}} &
\multicolumn{1}{c}{\parbox{0.3in}{\small Samp.\ error}} &
P \\ 
& & & & 
\multicolumn{1}{c}{km/s} & 
\multicolumn{1}{c}{} & 
\multicolumn{1}{c}{} &
\multicolumn{1}{c}{km/s} &
\multicolumn{1}{c}{km/s} &
\\ \hline 
LP      & BCG  & 119 & 8400  &  $832$ & $349$ & $51$ & 252 & 120 & 0.06 \\
SC      & TF   & 63  & 8100  &  $120$ & $295$ & $10$ & 140 & 170 & 0.30 \\
SMAC    & FP   & 56  & 6600  &  $690$ & $260$ & $-1$ & 200 & 180 & 0.29 \\
Willick & TF   & 15  & 11100 & $1060$ & $275$ & $28$ & 450 & 220 & 0.36 \\
EFAR    & FP   & 49  & 9500  &  $630$ & $53$  & $6$  & 380 & 290 & 0.16 \\ 
Tonry   & SNIa & 65  & 10300 &  $610$ & $311$ & $9$  & 200 & 130 & 0.58 \\
\hline 
{\bf STEWS$^{a}$}
& {\bf Mixed\/} & {\bfseries
248} & {\bfseries 8200} & $350$ & $288$ & $8$ & $80$ & 100 & \\ \hline 
\multicolumn{9}{l}{$^{a}$STEWS is S{\scriptsize MAC} + Tonry + 
E{\scriptsize FAR} + Willick + S{\scriptsize C}} \\ \hline 
\end{tabular}\\
\end{center}
\end{minipage}
\end{table}

In the penultimate column of Table 1, we present sampling errors for
the comparison between the bulk flow of the given survey compared to
an idealized dense and uniformly sampled sphere of radius 9000 \kms,
assuming a $\Lambda$CDM model with parameters: $\Omega_m = 0.35$,
$\Omega_{\Lambda} = 0.65$, $H_0 = 70$ km/s, $\Omega_b = 0.047$.  Note
that in nearly all cases the sampling errors are comparable to, or
larger than, the peculiar velocity errors. The LP and SNIa samples
have the highest density of objects and so have the smallest sampling
errors.  Because of its non-uniform sky coverage, the survey with the
highest sampling error is EFAR. The last column indicates the
probability that the bulk flow of a given sample is consistent, within
the errors, with the bulk flow from the other surveys. When sampling
errors are included there is {\em no conflict\/} for any survey at the
$2\sigma$ level. The sample in poorest agreement is the Lauer \&
Postman survey, but even there the difference is quite marginal
(significant at only the 94\% level).

We then throw all the data into the pot to cook. This process yields
the STEWS sample 
( = S{\scriptsize MAC} + Tonry + E{\scriptsize FAR} +
Willick + S{\scriptsize C}, 
but excluding LP) which has a bulk flow of
$350\pm80$ km/s toward \lb{288}{8}.  A plot of the peculiar velocities
for this sample is shown in Fig.\ 1.  For the $\Lambda$CDM model used
above, the expected rms value of the bulk flow, allowing for the
sparse geometry of the STEWS sample is 130 \kms\ in each component.
Allowing for random peculiar velocity errors, we find that the bulk
flow of the STEWS sample is consistent with the $\Lambda$CDM model.
The STEWS sample is obviously less sparse than the individual surveys
of which it is composed.  Sampling effects are still non-negligible,
however --- compared to an ideal densely-sampled survey of radius 9000
\kms, the sampling error is $\sim 100 \kms$.  Although the errors are
large, the bulk flow is still significantly different from zero.  It
thus appears that there are significant contributions to bulk motions
arising from scales $\gtrsim 100 \hmpc$.

\begin{figure}
\begin{center}
\psfig{figure=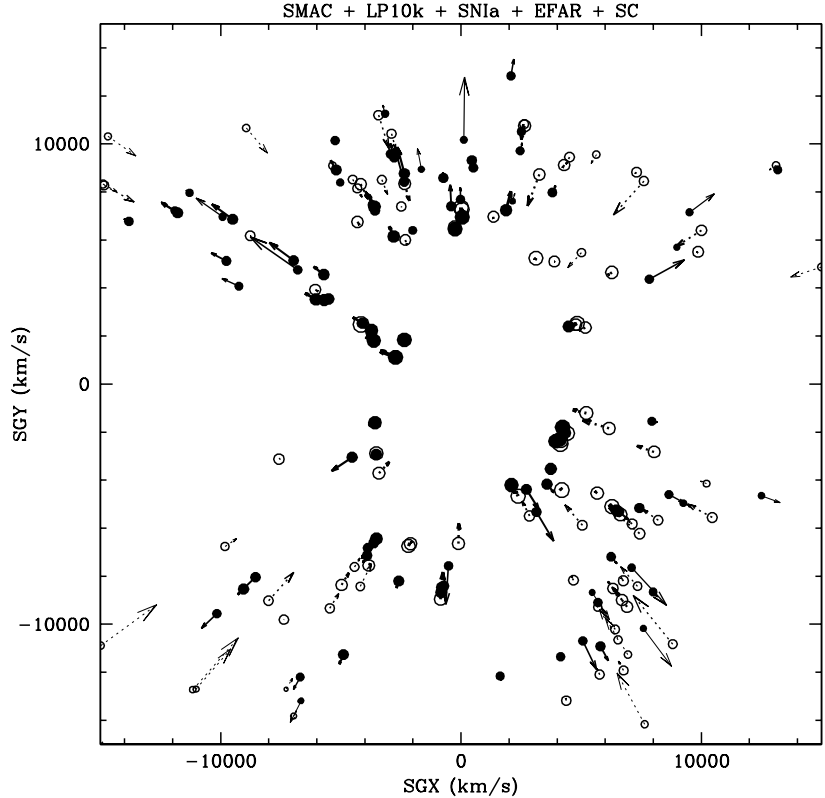,height=4.5in}
\end{center}
\caption{ Peculiar velocity diagram for the STEWS sample in the
Supergalactic Plane. The circle represents the distance to a cluster
and the tip of the vector represents its redshift.  Clusters with
smaller random errors, and hence greater statistical weight, are
indicated by larger circles. Outward flowing objects are solid,
inward-flowing ones are open with dotted vectors. Notice the excess of
inflowing objects on the right hand side, and the excess of outflowing
objects in the upper left quadrant.
}
\end{figure}

\section{Discussion}

These results suggest substantial contributions to the Local Group's
motion from large scales.  A natural question is whether we can
identify the structures responsible.  Part of the motion may be due to
the Shapley Concentration; the peculiar velocity data favor a
substantial mass for this supercluster complex.  If X-ray emitting
clusters are used to trace the gravity field, one predicts a very
strong contribution to the LG's motion from a distance of $\sim 150
\hmpc$ (Kocevski \emph{et al.}\,\cite{kem}). One prominent supercluster
at this distance in the right direction on the sky is the Shapley
Concentration.  On the other hand, Saunders and
collaborators\cite{sau} have extended the IRAS PSCz survey closer to
the Galactic Plane. They find that approximately 50\% of the Local
Group's motion, or $\sim 300 \kms$ in the direction of the negative
Galactic Y (\lb{270}{0}) arises from $\sim 200 \hmpc$. The
structure(s) responsible for this gravity have not been completely
identified.

We expect to resolve some of these issues with a new peculiar velocity
survey, the NOAO Fundamental Plane Survey (NFPS).  The NFPS is a
survey of 93 X-ray selected clusters within 200 \hmpc.  For each
cluster, we obtain 20 -- 70 FP distances per cluster, with a total of
4000 FP cluster galaxies.  The NFPS is therefore 4 times the size of
the SMAC and EFAR surveys combined.  We expect random, systematic and
sampling errors each $\lesssim 100 \kms$ and so expect to resolve the
question of the amplitude of the large-scale flow. A description of
this survey is given by Smith \emph{et al.}\,\cite{nfps}

\section{Summary}

We have compared the bulk flow of recent large-scale peculiar velocity
surveys to each other, allowing for the errors due to sparse
sampling. We conclude that, contrary to the current perception, there
is no significant conflict between these surveys. The combined STEWS
peculiar velocity dataset samples a volume $\sim 100 \hmpc$ in radius
and has a bulk flow of $350 \pm 80 \kms$.  Allowing for the sparse
sampling, we find that this result is not in conflict with the
$\Lambda$CDM models. Structure(s) responsible for the large-scale
motion have not yet been identified, but some likely suspects are
presently under surveillance.

\end{document}